\documentclass[fleqn,usenatbib]{mnras}

\usepackage{newtxtext,newtxmath}
\usepackage{hyperref}

\usepackage[T1]{fontenc}

\DeclareRobustCommand{\VAN}[3]{#2}
\let\VANthebibliography\thebibliography
\def\thebibliography{\DeclareRobustCommand{\VAN}[3]{##3}\VANthebibliography}

\usepackage{multicol}
\usepackage{caption}
\usepackage{subcaption}

\usepackage{graphicx}	
\usepackage{amsmath}	

\def\B{{\boldsymbol B}}
\def\B{{\boldsymbol B}}

\def\x{{\boldsymbol x}}
\def\y{{\boldsymbol y}}

\def\d{{\rm d}}

\title[Photospheric signatures of CME onset]{Photospheric signatures of CME onset}

\author[O.P.M. Aslam et al.]{
O.P.M. Aslam,$^{1}$
D. MacTaggart,$^{1}$\thanks{E-mail: david.mactaggart@glasgow.ac.uk}
T. Williams, $^{2}$
L. Fletcher$^{3,4}$
and
P. Romano$^{5}$
\\
$^{1}$School of Mathematics and Statistics, University of Glasgow, Glasgow, G12 8QQ, UK\\
$^{2}$Department of Mathematical Sciences, Durham University, Durham, UK\\
$^{3}$SUPA School of Physics and Astronomy, University of Glasgow, Glasgow G12 8QQ, UK\\
$^{4}$Rosseland Centre for Solar Physics, University of Oslo, PO Box 1029 Blindern, NO-0315 Oslo, Norway\\
$^{5}$INAF—Osservatorio Astrofisico di Catania, Via Santa Sofia 78, 95123 Catania, Italy
}

\date{Accepted 2024 September 09. Received 2024 September 06; in original form 2024 June 12}

\pubyear{2024}

\begin{document}
\label{firstpage}
\pagerange{\pageref{firstpage}--\pageref{lastpage}}
\maketitle

\begin{abstract}
Coronal mass ejections (CMEs) are solar eruptions that involve large-scale changes to the magnetic topology of an active region. There exists a range of models for CME onset which are based on twisted or sheared magnetic field above a polarity inversion line (PIL). We present observational evidence that topological changes at PILs, in the photosphere, form a key part of CME onset, as implied by many models. In particular, we study the onset of 30 CMEs and investigate topological changes in the photosphere by calculating the magnetic winding flux, using the \texttt{ARTop} code. By matching the times and locations of winding signatures with CME observations produced by the \texttt{ALMANAC} code, we confirm that these signatures are indeed associated with CMEs. Therefore, as well as presenting evidence that changes in magnetic topology at the photosphere are a common signature of CME onset, our approach also allows for the finding of the source location of a CME within an active region.

\end{abstract}




\section{Introduction}

The energy source of solar eruptions, such as flares and coronal mass ejections (CMEs), lies in the magnetic field \citep{forbes00}. As solar active regions emerge and evolve, the magnetic field forms configurations that can become unstable and lead to eruptions, and this scenario is the basis of many models and simulations of CME onset \citep[e.g.][]{antiochos99,amari00,roussev03,torok07,aulanier10,ishiguro17,jiang21}. A key element for the possibility of an eruption is that the magnetic field has a suitable \emph{magnetic topology}. This expression entails two important features. First, that the magnetic field has a suitable connectivity to allow for reconnection to occur efficiently and produce an eruption. Secondly, it describes how inherently twisted the magnetic field is. This latter point is related to \emph{magnetic helicity} which provides a lower bound for the free magnetic energy required to produce an eruption \citep[see][for empirical evidence]{tziotziou14,liokati23}.

Studies of magnetic topology have identified particular features that are closely associated with different eruptive events in the solar atmosphere. One such feature, referred to as a \emph{bald patch} \citep{titov93}, is a location where magnetic field lines become tangent to the photosphere. For example, bald patches have been associated with filaments \citep{lopez06}, brightenings \citep{fletcher01}, surges \citep{mandrini02} and flares \citep{lee21}. The importance of bald patches reflects the role they play in the two elements of magnetic topology outlined above. In terms of connectivity, two distinct bald patches can define separatrices, which are preferential regions for magnetic reconnection \citep{priest14}. In terms of twist, bald patches are to be expected along polarity inversion lines \citep{titov99}.

Bald patches are typically associated with concave up ``U-loops''. However, horizontal magnetic field at the photosphere can be indicative of more general topological behaviour, such as the emergence of a twisted or highly sheared magnetic field. In order to capture topological changes due to (near-)horizontal field at the photosphere, a suitable measure is \emph{magnetic winding} \citep{prior20,mactaggart21gafd,mactaggart21natcomms}. This quantity is a renormalization of magnetic helicity and provides a direct measure of field line topology. By removing the flux weighting from magnetic helicity, magnetic winding can be used to identify, in particular, topological complexity due to primarily near-horizontal field. Magnetic helicity, by contrast, is dominated by the strongest vertical field due to its flux weighting. For detailed discussions on the properties of magnetic winding, the reader is directed to the citations above. A brief summary of the results required for this work is presented in Appendix \ref{appendix}. In magnetograms, it is magnetic winding \emph{flux} $L$ that is calculated and this can indicate where and when there are large topological changes due, in the main, to changing near-horizontal magnetic field at the photosphere. 

The purpose of this work is two-fold. First, we present evidence that changes in magnetic topology at the photosphere are a common signature of CME onset. We analyze 30 CMEs and identify the topological onset signature at the photosphere by measuring the flux of magnetic winding. Secondly, and in parallel with the first point, we provide an approach for the identification of the location of the the origin of a CME within an active region, based solely on data from observations, i.e. without the explicit need to model the three-dimensional magnetic field.

The paper is set out as follows. First, the method of our approach for finding signatures of magnetic winding relating to CME onset is outlined in detail. Secondly, we present our results and discuss general features. We then present some example cases in more detail. The paper concludes with a summary and a discussion of the significance of our results.   

\section{Winding signature identification procedure}\label{sec:sigs}
In this section, we outline the steps involved in the identification of photospheric topological signatures of CME onset, which we will refer to as \emph{winding signatures}. Examples of outputs from each of the following parts will be presented later, as well as further descriptions of the analysis.

\subsection{Part 1: Time series calculations}
In this study, we present 30 CME events, making use of events already identified and studied in other works (see the CME list given in \cite{pal18} and the flare list given in \cite{liu12}). These events have been cross-checked to make sure that each CME occurred at a longitude (relative to the central meridian) within the range $(-60^\circ,+60^\circ)$. These limits are typical bounds used in the literature \citep[e.g.][]{park20} and represent the locations where winding (and helicity) calculations become compromised by projection effects. This is because our method of calculating the magnetic winding flux is based on Space-Weather Helioseismic and Magnetic Imager (HMI) Active Region Patches (SHARP) vector magnetograms \citep{hoeksema14}. The projection of magnetic field components onto a Cartesian grid is most accurate near the central meridian and the approximation begins to break down at the above limits. 

With the CME events selected, the first step in identifying the winding signature is to calculate a time series of $\dot{L} (={\rm d}L/{\rm d}t)$ using the \texttt{ARTop} code \citep{alielden23}. This code makes use of SHARP magnetograms, as mentioned above, utilising their full field of view, and the Differential Affine Velocity Estimator for Vector Magnetograms (DAVE4VM) technique \citep{schuck08}. For all calculations involving \texttt{ARTop} in this work, all magnetic field of strength lower than 20 Gauss is ignored, an apodizing window of 20 pixels is used for the DAVE4VM calculations and the full resolution of the magnetograms is used for the winding calculations.

For each of the 30 events, a time series starting from approximately 20 hours before the recorded coronagraph CME time (details are provided below) until a few hours after this time, is calculated. For events whose active regions are beyond the $-60^\circ$ longitude limit 20 hours before the recorded coronagraph CME time, the time series are calculated from when the active region has a longitude of $-60^\circ$.

For each time series, a running mean $\mu$ (based on the previous hour of data) is calculated and an envelope $\mu\pm2\sigma$, with a width of 2 standard deviations in both positive and negative directions, is determined. The \texttt{ARTop} code enables overlaying flare times and strengths on time series for a given active region. Where they occur, flare corresponding to CMEs are also recorded.

The peaks in magnetic winding, which we define as spikes of $\dot{L}$ penetrating the $\pm2\sigma$ envelope, near the recorded coronagraph time of the CME, are noted. The largest, in magnitude, of these spikes is taken as the initial marker for the winding signature. Typically, there will be one prominent peak shortly before the coronagraph time of CME and/or its corresponding flare. The analysis we are proposing here is, therefore, not primarily intended to be predictive, but rather a means of providing a deeper analysis of the birth of a given CME in the lower layers of the solar atmosphere. In general, the time series of a particular region also may contain other winding spikes that are associated with other topological events that are not directly related to CMEs, such as flares, emergence and motion in penumbrae (see later for more details). {A large spike in the magnitude of $\dot{L}$ corresponds to a large amount of topologically-significant (i.e. twisted and sheared) magnetic field passing through the photosphere within a given time. This, as we will demonstrate, typically occurs in a localized part of the active region on or near a PIL.}

\subsection{Part 2: Magnetic winding maps - winding signature time and location}
At each time, the quantity $\dot{L}$ is the spatial integral over the map of the field line winding flux $\dot{\mathcal{L}}$ (see the Appendix for further details). By examining the maps of $\dot{\mathcal{L}}$, at the times near the peaks recorded in Part 1, it is possible to determine, {by visual inspection}, for each CME, the location within active region where the winding signature occurs. The time of the winding signature corresponds to when the structure corresponding to the largest spike first appears in the maps of $\dot{\mathcal{L}}$. This is normally co-temporal with the main spike or occurs shortly before ($\sim$12-24 mins) the main spike. 

Proposed mechanisms for CME onset (as in the examples cited previously) are based on sheared or twisted magnetic field above the main PIL of the active region. Thus, time series peaks identified in Part 1 must correspond with signatures at PILs in order to be considered to be related to CME onset (see \cite{moratis24} for a related study on the connection between flares and relative field line helicity at PILs). {The strength of the winding signature depends on the particular eruption mechanism that takes place. Later, we will present different pre-eruption magnetic topologies that can lead to strong winding signatures through topological changes at the photosphere.} 

\subsection{Part 3: Confirmation with ALMANAC}


In order to link the photospheric winding signatures with CME onset, the final stage is to connect their times and locations with the earliest available observations of the CMEs higher in the atmosphere (with respect to the photosphere). Although we base our selection of 30 events on previously recorded CMEs, as detailed below, these use observations in the high atmosphere (from coronagraphs) and, therefore, do not represent the early stage of CME onset. Instead, we need information from heights lower than those available from coronagraph data.

To perform this task systematically, we make use of the \texttt{A}utomated Detection of Corona\texttt{L} \texttt{MA}ss Ejecta origi\texttt{N}s for Space Weather \texttt{A}ppli\texttt{C}ations (\texttt{ALMANAC}) code \citep{williams22}. \texttt{ALMANAC}, unlike many widely adopted CME detection methods does not rely upon coronagraph data, but instead utilises data from the Atmospheric Imaging Assembly \citep[AIA]{lemen12}. The main advantage of \texttt{ALMANAC} is that it does not require geometrical fitting to approximate the CME source location in the low solar corona. Thus, the code does not inherently have large uncertainties due to projection effects caused by fitting a simple ``wire-frame'' of a three-dimensional object mapped in two dimensions. As such, \texttt{ALMANAC} provides a reliable (low-corona) CME origin that is obtained independently of any winding signatures from the earlier phases of an eruption.

To detect potential Earth-directed CMEs, \texttt{ALMANAC} first crops the map size to eliminate off-limb contributions and standardises the intensity across an 8-hour image sequence by thresholding intensities and normalising the data values. It is then smoothed through convolution  and subtracted from the normalised data to create a high-bandpass and time-filtered image sequence. Each time step of the time-filtered data is then divided by the median of the absolute values of the unfiltered data to eliminate contribution from ``static'' structures such as active regions. The method employs a series of Boolean masks to isolate connected clusters of pixels associated with a potential eruption, and spatio-temporal smoothing of these masks helps avoid the segmentation of regions. The first time step in which a region of sufficient size and duration is identified is used as the CME onset time, whilst the center of mass for the masked pixels at that time provides the central location for the CME. Full details can be found in \citet{williams22}.


\section{Summary of the data}
We now present results for the 30 selected events. Details are displayed in Table \ref{table}.

\begin{table*}
	\centering
	\caption{A table of 30 CME events with winding signatures related to CME onset. The columns are, in order, the active region NOAA number, the time of the winding signature, the CME time from \texttt{ALMANAC},  the recorded CME time (LASCO or STEREO), the winding signature coordinates in Carrington longitude and sine latitude (to the nearest degree), the CME location from \texttt{ALMANAC} in Carrington longitude and sine latitude (to the nearest degree) and the relative difference of the winding (\texttt{ARTop}) and \texttt{ALMANAC} longitudes and latitudes divided by the longitudinal and latitudinal extensions of the active region respectively (rounded to three decimal places). The Carrington longitudes are normalized to lie within the range [$0^\circ,360^\circ$].}
	\label{table}
	\begin{tabular}{lllllllll} 
		\hline
		Active region & Winding obs. time & \texttt{ALMANAC} obs. time & Recorded CME obs. time  & Winding coord.  &\texttt{ALMANAC} coord. & $\Delta_{\rm Lon.}$ & $\Delta_{\rm Lat.}$\\
        (NOAA) & &  & (LASCO or STEREO)  & (lon.,lat.) & (lon.,lat.) & &\\
		\hline
11081 &	2010 Jun 11, 22:48:00 & 2010 Jun 12, 00:30:08 & 	 2010 Jun 12, 01:31:39 &	 	 (103$^\circ$, 23$^\circ$) & (104$^\circ$, 23$^\circ$) & 0.125 & 0\\
11158 &	2011 Feb 14, 16:24:00 &2011 Feb 14, 17:50:00 &	 2011 Feb 14, 18:24:06 &	 	  (29$^\circ$, -22$^\circ$) & (24$^\circ$, -16$^\circ$) & 0.294 & 0.75\\
11158 &	2011 Feb 15, 01:24:00 &2011 Feb 15, 01:40:00 &	 2011 Feb 15, 02:24:05 &	 	  (36$^\circ$, -19$^\circ$) & (36$^\circ$, -17$^\circ$) & 0 & 0.222\\
11162 &	2011 Feb 18, 08:48:00 &2011 Feb 18, 09:00:08 &	 2011 Feb 18, 11:45:00 &	 	  (337$^\circ$, 21$^\circ$) & (339$^\circ$, 21$^\circ$) & 0.2 & 0 \\
11226 & 2011 Jun 01, 16:00:00 &2011 Jun 01, 16:20:00 & 	2011 Jun 01, 18:36:00 &	 	 (34$^\circ$, -23$^\circ$) & (39$^\circ$, -16$^\circ$) & 0.294 & 0.7\\
11227 & 2011 Jun 02, 06:00:00 &2011 Jun 02, 06:20:00 & 	2011 Jun 02, 08:12:06 &	 	 (27$^\circ$, -18$^\circ$) & (25$^\circ$, -15$^\circ$) & 0.1 & 0.3\\
11247 & 2011 Jul 08, 23:24:00 &2011 Jul 08, 23:40:08 &  2011 Jul 09, 00:48:05 &	 	 (269$^\circ$, -19$^\circ$) & (254$^\circ$, -31$^\circ$) & 0.75 & 2\\
11261 & 2011 Aug 03, 12:48:00 &2011 Aug 03, 13:10:08 &	2011 Aug 03, 14:00:07 &	 	  (335$^\circ$, 17$^\circ$) & (334$^\circ$, 24$^\circ$) & 0.038 & 0.438\\
11261 & 2011 Aug 04, 03:00:00 &2011 Aug 04, 03:20:09 &	2011 Aug 04, 04:12:05 &	 	  (330$^\circ$, 17$^\circ$) & (331$^\circ$, 21$^\circ$) & 0.033 & 0.235\\
11283 & 2011 Sep 06, 21:36:00 &2011 Sep 06, 21:50:08 &	2011 Sep 06, 23:05:57 &	 	  (223$^\circ$, 14$^\circ$) & (213$^\circ$, 14$^\circ$) & 0.5 & 0\\
11283 & 2011 Sep 07, 21:24:00 &2011 Sep 07, 22:10:09 &	2011 Sep 07, 23:05:58 &	 	  (228$^\circ$, 15$^\circ$) & (231$^\circ$, 18$^\circ$) & 0.15 & 0.188\\
11302 & 2011 Sep 24, 18:36:00 &2011 Sep 24, 18:50:09 &	2011 Sep 24, 19:36:06 &	 	  (286$^\circ$, 13$^\circ$) & (291$^\circ$, 12$^\circ$) & 0.217 & 0.063\\
11318 & 2011 Oct 15, 03:00:00 &2011 Oct 15, 04:10:09 &	2011 Oct 15, 05:12:05 &	 	  (97$^\circ$, 20$^\circ$) & (100$^\circ$, 23$^\circ$) & 0.273 & 0.5\\
11384 & 2011 Dec 26, 10:00:00 &2011 Dec 26, 10:30:08 &	2011 Dec 26, 11:48:07 &	 	  (200$^\circ$, 10$^\circ$) & (208$^\circ$, 4$^\circ$) & 0.25 & 0.231\\
11422 & 2012 Feb 19, 08:00:00& 2012 Feb 19, 08:20:02 &	2012 Feb 19, 09:36:06 &	 	  (177$^\circ$, 16$^\circ$) & (173$^\circ$, 19$^\circ$) & 0.4 & 0.375\\
11466 & 2012 Apr 27, 07:24:00& 2012 Apr 27, 08:10:08 &	2012 Apr 27, 10:49:12 &	 	  (40$^\circ$, 14$^\circ$) & (45$^\circ$, 11$^\circ$) & 0.417 & 0.375\\
11577 & 2012 Sep 27, 22:00:00 &2012 Sep 27, 23:10:08 &	2012 Sep 28, 00:12:05 &	 	  (166$^\circ$, 9$^\circ$) & (172$^\circ$, 8$^\circ$) & 0.2 & 0.067\\
11618 & 2012 Nov 20, 19:00:00 &2012 Nov 20, 19:10:08 &	2012 Nov 20, 20:05:00 &	 	  (135$^\circ$, 8$^\circ$) & (136$^\circ$, 6$^\circ$) & 0.033 & 0.2 \\
11618 & 2012 Nov 21, 06:24:00 &2012 Nov 21, 06:30:07 &	2012 Nov 21, 07:25:00 &	 	  (132$^\circ$, 6$^\circ$) & (136$^\circ$, 5$^\circ$) & 0.133 & 0.1\\
11776 & 2013 Jun 18, 23:36:00 &2013 Jun 19, 00:50:07 & 	2013 Jun 19, 01:25:50 &	 	 (252$^\circ$, 11$^\circ$) & (255$^\circ$, 8$^\circ$) & 0.375 & 0.5\\
11810 & 2013 Aug 07, 15:00:00 &2013 Aug 07, 15:00:07 &	2013 Aug 07, 18:24:05 &	 	  (325$^\circ$, -30$^\circ$) & (328$^\circ$, -37$^\circ$) & 0.136 & 0.636\\
11865 & 2013 Oct 13, 00:00:00 &2013 Oct 13, 00:10:08 &	2013 Oct 13, 00:51:00 &	 	  (144$^\circ$, -21$^\circ$) & (145$^\circ$, -29$^\circ$) & 0.04 & 0.533\\
11870 & 2013 Oct 16, 13:36:00 &2013 Oct 16, 14:50:07 &	2013 Oct 16, 15:48:05 &	 	  (143$^\circ$, -13$^\circ$) & (149$^\circ$, -13$^\circ$) & 0.2 & 0\\
11875 & 2013 Oct 22, 13:48:00 &----- 		          &	2013 Oct 22, 15:15:00 &	 	  (29$^\circ$, 6$^\circ$) & ----- & ----- & -----\\
11875 & 2013 Oct 22, 20:12:00 &2013 Oct 22, 20:40:08 &	2013 Oct 22, 21:55:00 &	 	  (30$^\circ$, 5$^\circ$) & (35$^\circ$, 4$^\circ$) & 0.25 & 0.059\\
11891 & 2013 Nov 08, 08:12:00 &2013 Nov 08, 09:20:07 &	2013 Nov 08, 11:12:05 &	 	  (204$^\circ$, -18$^\circ$) & (207$^\circ$, -19$^\circ$) & 0.3 & 0.125\\
11946 & 2014 Jan 07, 02:12:00 &2014 Jan 07, 02:20:07 & 	2014 Jan 07, 03:36:05 &	 	 (101$^\circ$, 9$^\circ$) & (104$^\circ$, 11$^\circ$) & 0.231 & 0.25\\
12089 & 2014 Jun 10, 23:12:00 &2014 Jun 11, 00:10:09 & 	2014 Jun 11, 00:48:06 &	 	 (196$^\circ$, 17$^\circ$) & (195$^\circ$, 17$^\circ$) & 0.1 & 0\\
12230 & 2014 Dec 09, 08:48:00 &2014 Dec 09, 09:20:01 &	2014 Dec 09, 10:24:05 &	 	  (319$^\circ$, -16$^\circ$) & (320$^\circ$, -17$^\circ$) & 0.05 & 0.083\\
12465 & 2015 Dec 11, 15:48:00  &2015 Dec 11, 16:50:06 &	2015 Dec 11, 18:24:04 &	 	  (168$^\circ$, -4$^\circ$) & (171$^\circ$, -6$^\circ$) & 0.176 & 0.25\\
		\hline
	\end{tabular}
\end{table*}

We argue that these data provide strong evidence that the winding signatures at the photosphere are related to CME onset due to the close association of the times and locations of winding signatures to both early observations of CMEs and their associated flares.


There is a strong match between the locations of the winding signatures and the CME positions determined by \texttt{ALMANAC} for all events with the exception of the CME in AR11247, for which the difference in both longitude and latitude is greater than 10$^\circ$; and the first event of AR11875, for which the CME was not detected by \texttt{ALMANAC}. We will discuss the event of AR11247 in more detail later. 

Differences in position are typically only a few degrees, with the largest separation (not including the event of AR11247) being 10$^\circ$ (for the Carrington longitude of the first event of AR11283). With the exception of the event of AR11247 and the first event of AR11875, the differences in latitude and longitude between the winding measurements from \texttt{ARTop} and the CME locations from \texttt{ALMANAC} are less than the active region dimensions at the photosphere. This is shown in Table \ref{table} by the $\Delta$-quantities, which measure
\[
\Delta_X = \left|\frac{X_{\texttt{ARTop}} -X_\texttt{ALMANAC}}{X_{\rm AR}}\right|,
\]
where $X$ represents longitude or latitude and $X_{\rm AR}$ represents the longitudinal or latitudinal extent of the active region. Some deflection is expected for CMEs due to the nearby constraints of surrounding and overlying magnetic field, effects of the rise mechanism \citep[such as twisting, e.g.][]{vourlidas11} and the heliospheric current sheet \citep{kay17}. However, as indicated by the $\Delta$-quantities in Table \ref{table}, {since all these values are less than 1}, such deflections are generally still within the photospheric area of the active region (at least for the early onset times presented here). {In addition to the physical grounds for deflection, listed above, such a close matching between the locations produced by \texttt{ARTop} and \texttt{ALMANAC} is not necessarily guaranteed since the latter can be influenced by surrounding regions and strong flares. \cite{williams22} report that for the twenty halo CMEs that they study with \texttt{ALMANAC}, CME locations are within $10^\circ\pm10^\circ$ of X-ray emission values reported by RHESSI for
that sample - a typically much larger margin than that which we find based on $\Delta<1$.}

The first times listed in Table \ref{table} are those of the winding signatures, the determination of which was outlined in the previous section. These times are constrained by \texttt{ARTop}'s magnetogram cadence, which is 12 minutes (i.e. that of the SHARP magnetograms). The next column of times is that of when the \texttt{ALMANAC} code detects a CME from AIA data (constrained to a 10-minute cadence). The last column of times are from existing catalogues of CME observations, in particular those made using the Large Angle and Spectrometric Coronagraph (LASCO) onboard the Solar and Heliospheric Observatory (SOHO) mission \citep{brueckner95} and the Solar TErrestrial RElations Observatory (STEREO) \citep{kaiser08}. As mentioned earlier, it is these recorded coronagraph times that were used to select CMEs and help identify the corresponding winding signatures.

These times, following the discussion earlier, correspond to observations of different heights in the atmosphere: winding - photosphere, \texttt{ALMANAC} - low atmosphere, coronagraph times - high atmosphere. Therefore, if the winding signatures correspond to CME onset, rather than some later phase of CME evolution, the winding times should be closer to the \texttt{ALMANAC} times than the coronagraph times. This is indeed the case, with all of the winding times preceding the \texttt{ALAMANC} times. Those events for which the winding time is less than 12 minutes (the \texttt{ARTop} cadence) before the \texttt{ALMANAC} time, may be considered to be approximately co-temporal, due to the available time resolution. As mentioned previously, the first CME of AR11875 was not detected with \texttt{ALMANAC} and so we cannot make as clear a connection between the observations at the photosphere and those higher in the atmosphere for this case.

Since CMEs are normally, though not always, associated with flares, we now compare the winding signature times to the start times of the flares that are associated with CMEs, i.e. those that are closest in time to the recorded coronograph CME times. Table \ref{table_flare} lists the flares associated with each of the 30 events and shows how their start times compare with the winding signature times.

\begin{table*}
	\centering
	\caption{The corresponding flares of the 30 CME events. For each event, the GOES flare strength, the winding observation time and the flare start time is shown.}
	\label{table_flare}
	\begin{tabular}{llll} 
		\hline
		Active region & Flare strength & Winding obs. time & Flare start time \\
        (NOAA) & (GOES) & & (NOAA)  \\
		\hline
11081    &	M2.0&	2010 Jun 11, 22:48 &	2010 Jun 12, 00:30		\\
11158    &	M2.2&	2011 Feb 14, 16:24&	2011 Feb 14, 17:20		\\
11158    &	X2.2&	2011 Feb 15, 01:24&	2011 Feb 15, 01:44		\\
11162    &	M1.0&	2011 Feb 18, 08:48&	2011 Feb 18, 10:23		\\
11226    &	C4.1&	2011 Jun 01, 16:00&	2011 Jun 01, 16:51		\\
11227    &	C3.7&	2011 Jun 02, 06:00&	2011 Jun 02, 07:22		\\
11247    &	B4.7&	2011 Jul 08, 23:24&	2011 Jul 08, 23:57		\\
11261    &	M6.0&	2011 Aug 03, 12:48&	2011 Aug 03, 13:17		\\
11261    &	M9.7&	2011 Aug 04, 03:00&	2011 Aug 04, 03:41		\\
11283   &	X2.1&	2011 Sep 06, 21:36&	2011 Sep 06, 22:12		\\
11283   &	X1.8&	2011 Sep 07, 21:24&	2011 Sep 07, 22:32		\\
11302   &	M3.0&	2011 Sep 24, 18:36&	2011 Sep 24, 19:09		\\
11318   &	C2.3&	2011 Oct 15, 03:00&	2011 Oct 15, 04:19		\\
11384   &	C5.7&	2011 Dec 26, 10:00&	2011 Dec 26, 11:23		\\
11422   &	C1.0&	2012 Feb 19, 08:00&	2012 Feb 19, 08:41		\\
11466   &	M1.0&	2012 Apr 27, 07:24&	2012 Apr 27, 08:15		\\
11577   &	C3.7&	2012 Sep 27, 22:00&	2012 Sep 27, 23:36		\\
11618   &	M1.6&	2012 Nov 20, 19:00&	2012 Nov 20, 19:21		\\
11618   &	M1.4&	2012 Nov 21, 06:24&	2012 Nov 21, 06:45		\\
11776   &	C2.3&	2013 Jun 18, 23:36&	2013 Jun 19, 00:50		\\
11810   &	B4.8&	2013 Aug 07, 15:00&	2013 Aug 07, 14:42		\\
11865   &	M1.7&	2013 Oct 13, 00:00&	2013 Oct 13, 00:12		\\
11870   &	C1.8&	2013 Oct 16, 13:36&	2013 Oct 16, 15:03		\\
11875   &	M1.0&	2013 Oct 22, 13:48&	2013 Oct 22, 14:49		\\
11875   &	M4.2&	2013 Oct 22, 20:12&	2013 Oct 22, 21:15		\\
11891   &	M2.3&	2013 Nov 08, 08:12&	2013 Nov 08, 09:22		\\
11946   &	M1.0&	2014 Jan 07, 02:12&	2014 Jan 07, 03:49	 	\\
12089   &	C2.1&	2014 Jun 10, 23:12&	2014 Jun 10, 23:46		\\
12230   &	C8.6&	2014 Dec 09, 08:48&	2014 Dec 09, 09:58		\\
12465   &	C5.6&	2015 Dec 11, 15:48&	2015 Dec 11, 16:48		\\
		\hline
	\end{tabular}
\end{table*}

The winding signatures precede or are approximately co-temporal with the flare start times for all events except that of AR11810. This event and that of AR11247 are the only ones for which the associated flare is B-class. For C-class and above, there is a consistent picture, providing further evidence that the winding signatures are related to CME onset. It is worth mentioning that since the majority of the winding signatures precede flares, for the events studied, {we can exclude the possibility that these signatures are influenced by any instrumental effects related to the inversion of the Fe I line at 617.3 nm due to the strong flare emission.}

\section{Analysis of example regions}
We now demonstrate the nature of the winding signature for three active regions from Table \ref{table}, with different morphologies, connecting it to the global magnetic topology of the active region. For each example region, we display the outputs (time series and maps) from \texttt{ARTop}, from which the winding signature is identified. In order to connect this signature to the global region topology, we approximate the three-dimensional magnetic field at the time of the winding signature by a nonlinear force-free field. We make use of the deep learning method of \cite{jarolim23} which produces a solution to the equations
\begin{align}
    (\nabla\times\B)\times\B &= \boldsymbol{0},\label{ff1} \\
    \nabla\cdot\B &= 0, \label{ff2}
\end{align}
where $\B$ is the magnetic field vector, extrapolating from the SHARP magnetograms of the given region. Clearly, satisfying equation (\ref{ff1}) is an approximation (a static representation of a dynamic process) but many studies indicate that force-free extrapolations provide a good representation of the global topology of an active region magnetic field. 

{To test the accuracy of our extrapolations, we consider the metrics used by \cite{wheatland00} and \cite{jarolim23}. As a measure of the solenoidality, which indicates how well an extrapolation represents a magnetic field, we calculate the $\B$-weighted divergence averaged over the entire domain, and refer to this quantity as $\lambda$,
\[
\lambda = \left\langle\frac{\|\nabla\cdot\B\|_{2}}{\|\B\|_{2}}\right\rangle. 
\]
As a measure of force-freeness, we first calculate, at every grid point $i$, the sine of the angle between the current density and the magnetic field. Writing $F_i=\|(\nabla\times\B)\times\B\|_{2}$, $B_i = \|\B\|_{2}$ and $J_i = \|\nabla\times\B\|_{2}$ for the norms evaluated at grid point $i$
,\[
\sigma_i = \sin\theta_i = \frac{F_i}{B_iJ_i}.
\]
From this, a current-weighted measure of the angle is produced,
\[
\theta_J = \arcsin\left(\frac{{\sum_i}J_i\sigma_i}{\Sigma_iJ_i}\right).
\]
The values of $\lambda$ and $\theta_J$ are reported for each extrapolation. We note that the extrapolations presented here are solely for the purpose  of giving indications of the magnetic topology within active regions and are not used as the basis of any further topological calculations, as in, for example, \cite{valori16} and \cite{thalmann20}.

}

\subsection{AR11318}
\begin{figure*}
    \centering
    \includegraphics[scale=0.4]{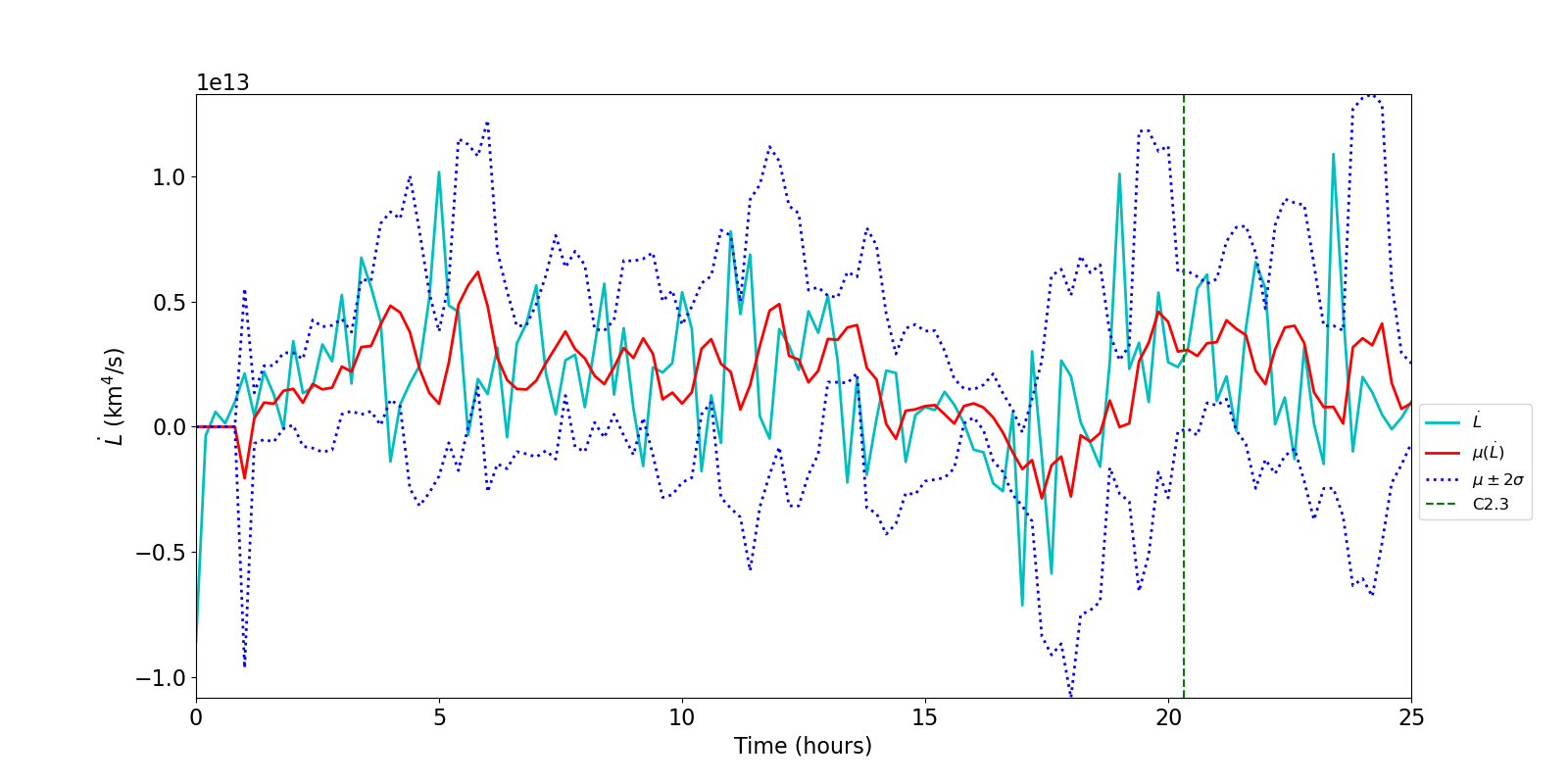}
    \caption{Time series of the magnetic winding flux rate $\dot{L}$ before the CME of AR11318. The quantities displayed follow the description in Part 1 of Section \ref{sec:sigs}. The LASCO CME time is at 21.2 hours on this graph.}
    \label{AR11318_time_series}
\end{figure*}

The first example region we consider, AR11318, has been studied in several other works \citep{romano14,mactaggart21natcomms}. Compared to other active regions, it is relatively simple - it is a bipolar region in which one flare and one CME occurred. Figure \ref{AR11318_time_series} displays the time series of $\dot{L}$ in the period before the CME.

\begin{figure}
    \centering
    \begin{subfigure}{\columnwidth}
        \includegraphics[width=\textwidth]{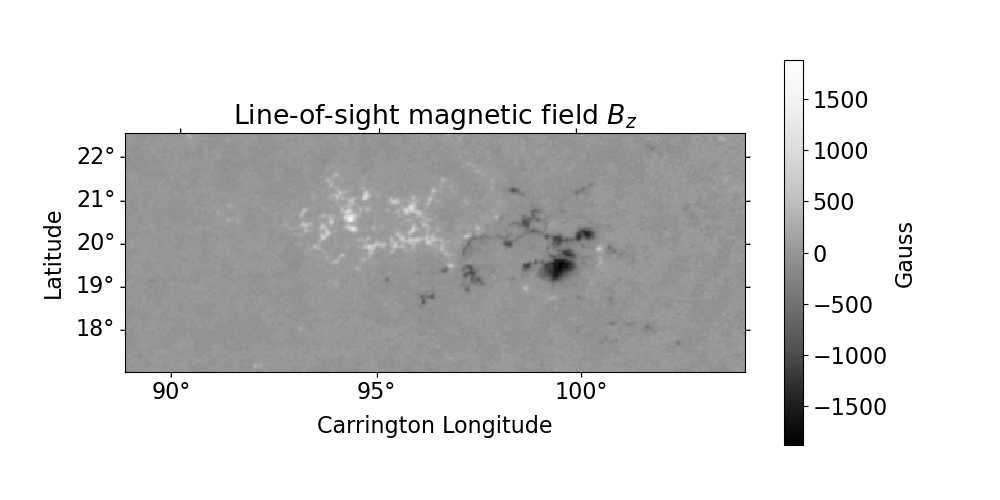}
        \caption{}
        \end{subfigure}

    \begin{subfigure}{\columnwidth}
        \includegraphics[width=\textwidth]{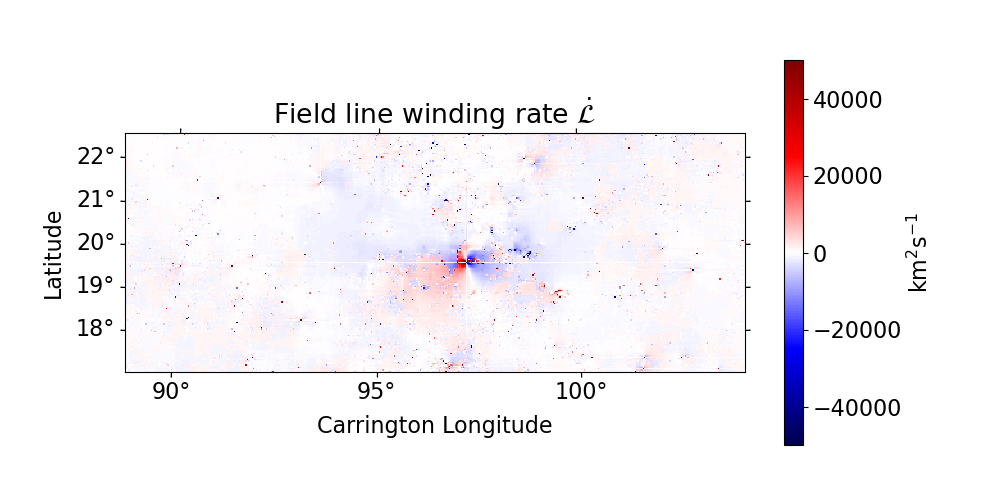}
        \caption{}
      \end{subfigure}

      \begin{subfigure}{\columnwidth}
        \includegraphics[width=\textwidth]{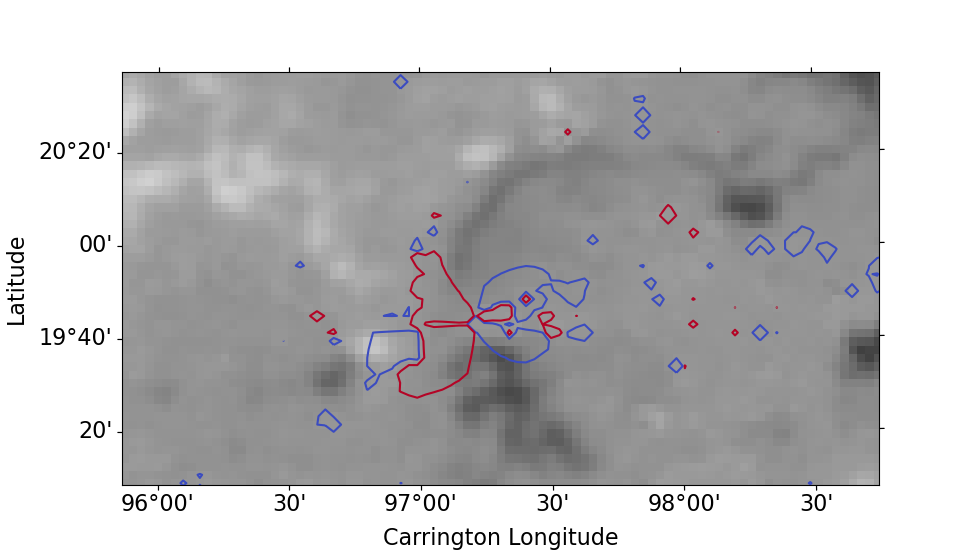}
        \caption{}
      \end{subfigure}
    \caption{(a) shows a map of $B_z$, (b) a map of $\dot{\mathcal{L}}$ and (c) a detail of the $B_z$ map with contours ($\pm10000$ km$^2$s$^{-1}$) of $\dot{\mathcal{L}}$ highlighting the centre of the winding signature at the base of the central curving PIL. All maps correspond to the time of the winding spike identified in Figure \ref{AR11318_time_series}.}\label{AR11318_maps}
\end{figure}

On the time scale of Figure \ref{AR11318_time_series}, there is a C2.3 flare beginning at 20.32 hours, followed shortly by the first sighting of a CME by LASCO at 21.2 hours. The \texttt{ALMANAC} time for the CME is approximately co-temporal with the C2.3 flare (a 9-minute difference). There is a particularly large spike in $\dot{L}$ at 19 hours and this, by the method outlined in Section \ref{sec:sigs}, is taken as the winding signature (it is the only spike near the flare and LASCO CME times that penetrates the $2\sigma$ envelope). The location of this signature within AR11318 can be found by considering the maps in Figure \ref{AR11318_maps}.

There is a rapid change in the field line winding $\mathcal{L}$ at one particular location, as seen in Figure \ref{AR11318_maps} (b). Comparing this with the magnetogram in Figure \ref{AR11318_maps} (a), this location corresponds to a small PIL between the two main spots{, curving downward at about 97$^\circ$ Carringtion longitude. Figure \ref{AR11318_maps} (c) displays a zoomed-in window of the the $B_z$ map at the location of the winding signature, indicated by contours of $\dot{\mathcal{L}}$ ($\pm$10000 km$^2$s$^{-1}$).}

\begin{figure}
    \centering
    \includegraphics[width=0.9\linewidth]{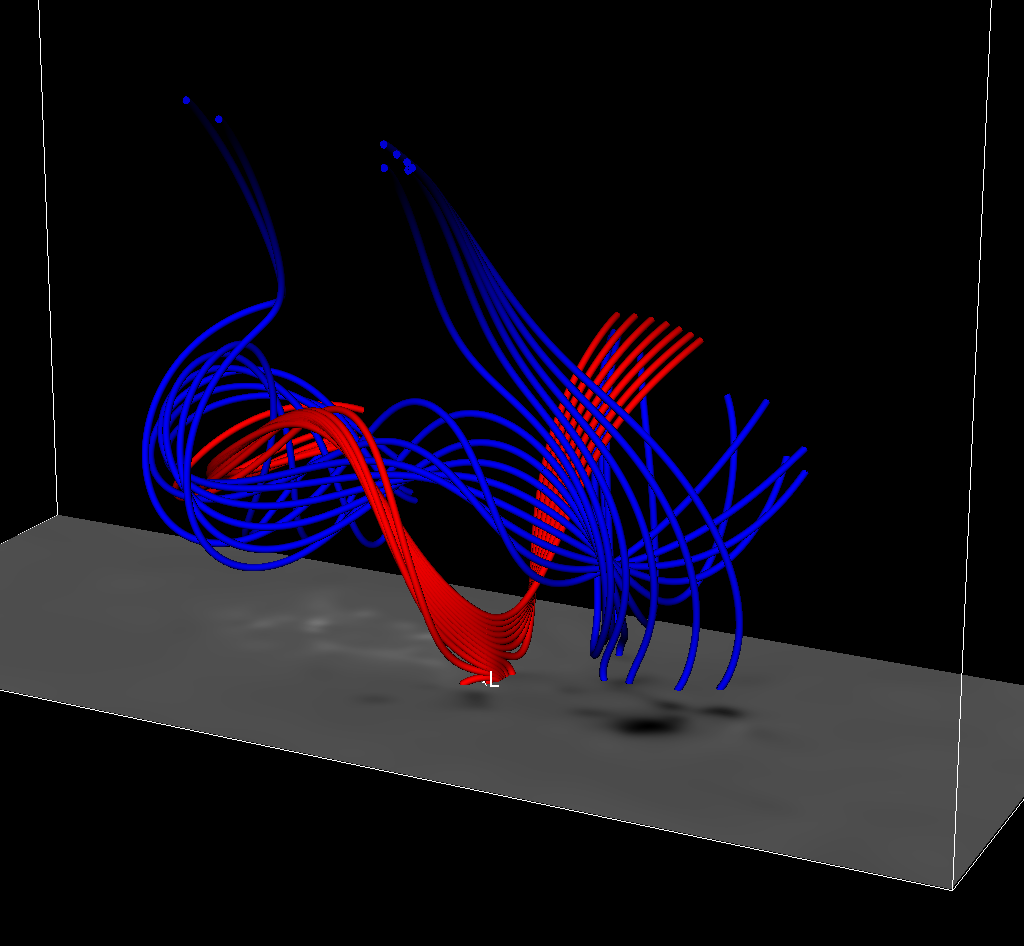}
    \caption{A nonlinear force-free extrapolation of the field lines of AR11318 at the time of the winding signature. The bald patch field lines, shown in red, are traced from the location of the winding signature, {indicated by ``L''. Surrounding field lines, shown in blue, trace out a flux rope structure. The metrics for this extrapolation are $\lambda=0.0018$ and $\theta_J=2.73^\circ$.}}
    \label{AR11318_field_lines}
\end{figure}

To better understand what this feature is, we consider a nonlinear force-free extrapolation, shown in Figure \ref{AR11318_field_lines}.

The red field lines, traced from the winding signature location, {indicated by ``L'',} reveal a bald patch which, as described in the Introduction, is a common topological feature related to eruptive phenomena. The winding signature here could refer to the emergence or submergence of the sheared bald patch field lines. Another possible explanation is due to reconnection related to the rising CME flux rope. The extrapolation traces out (shown in blue in Figure \ref{AR11318_field_lines}), a sigmoidal flux rope, which matches well the shape of features observed in AIA 193 \AA~and 304 \AA~ \citep{romano14}.

\subsection{AR11158}
\begin{figure*}
    \centering
    \includegraphics[scale=0.4]{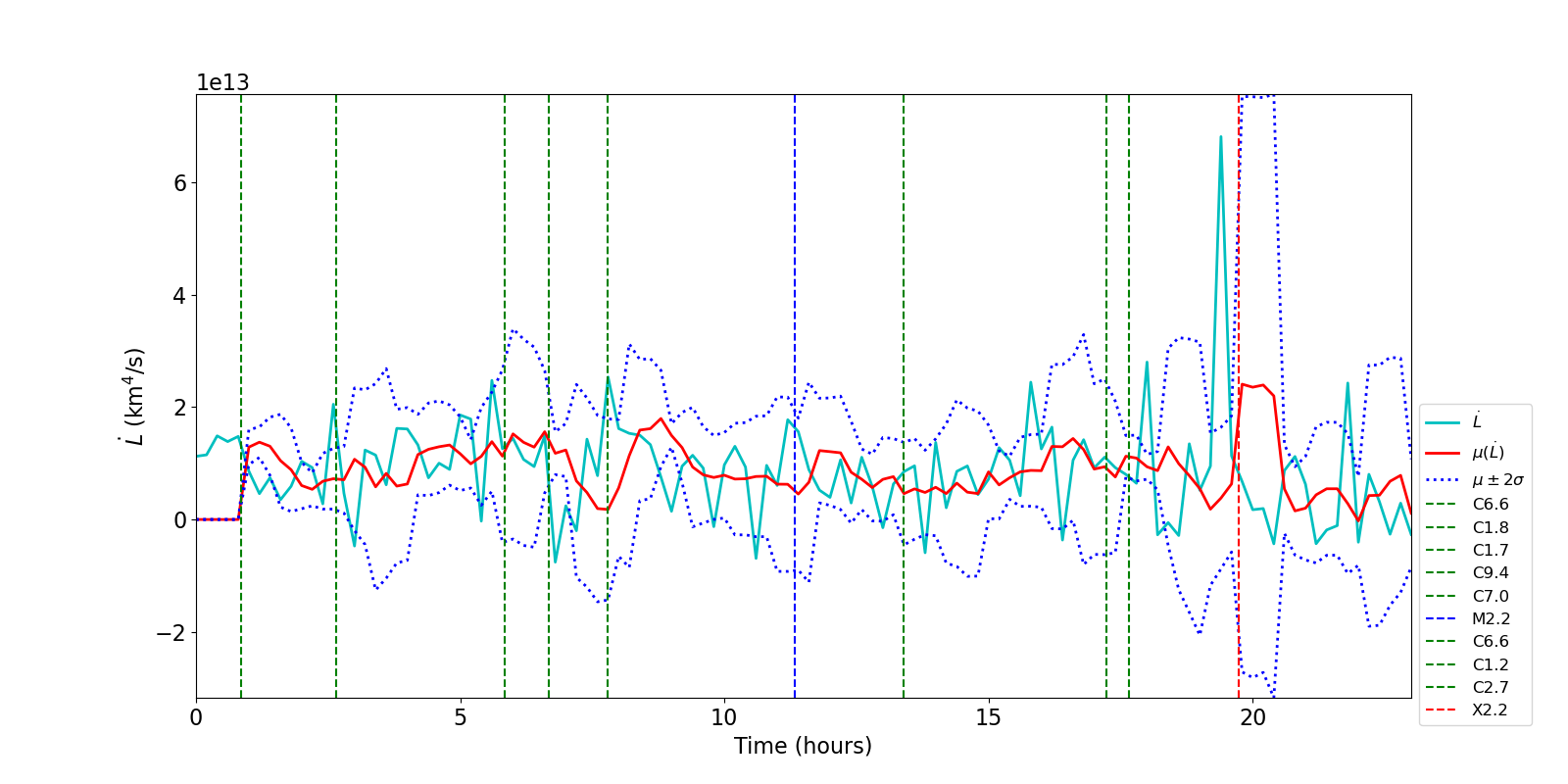}
    \caption{Time series of the magnetic winding flux rate $\dot{L}$ before the chosen CME of AR11158. The quantities displayed follow the description in Part 1 of Section \ref{sec:sigs}. The LASCO CME time is at 20.4 hours on this graph.}
    \label{AR11158_time_series}
\end{figure*}

\begin{figure}
    \begin{subfigure}[]{\columnwidth}
        \includegraphics[width=\textwidth]{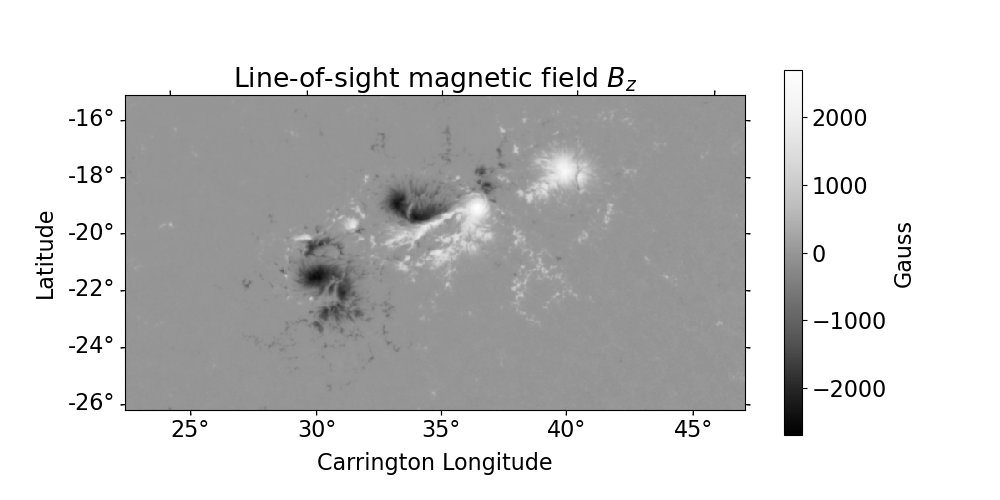}
        \caption{}
        \end{subfigure}

    \begin{subfigure}[]{\columnwidth}
        \includegraphics[width=\textwidth]{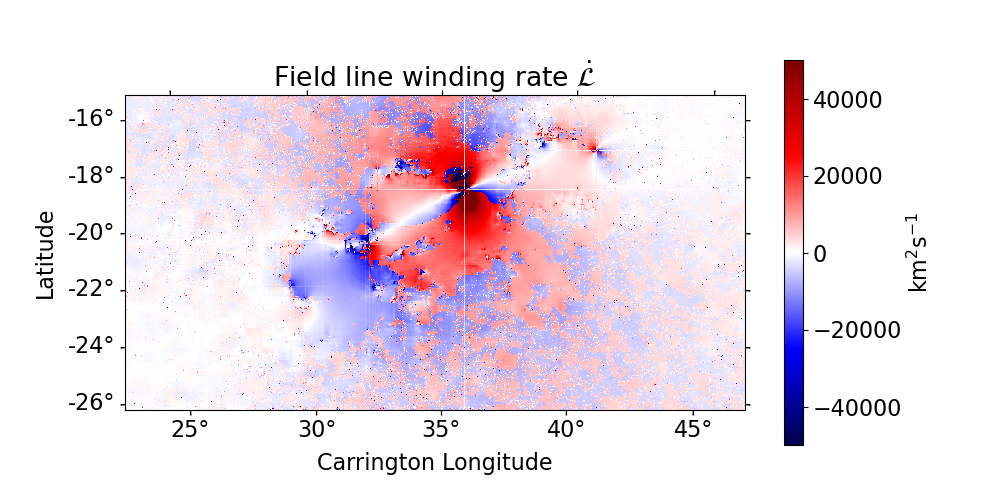}
        \caption{}
      \end{subfigure}
      
      \begin{subfigure}[]{\columnwidth}
        \includegraphics[width=\textwidth]{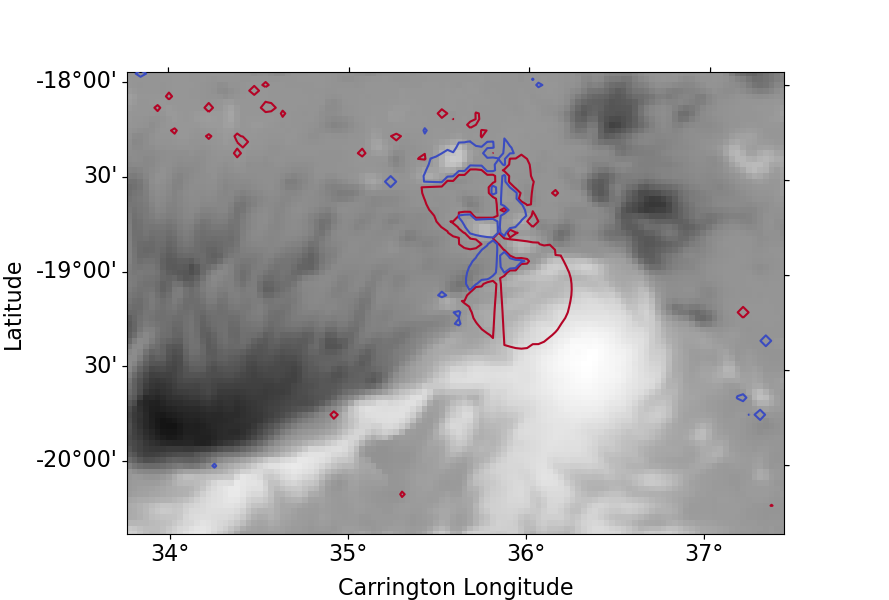}
        \caption{}
      \end{subfigure}
      
    \caption{(a) shows a map of $B_z$, (b) a map of $\dot{\mathcal{L}}$ and (c) a detail of the $B_z$ map with contours ($\pm60000$ km$^2$s$^{-1}$) of $\dot{\mathcal{L}}$ highlighting the centre of the winding signature at the top of the central PIL. All maps correspond to the time of the winding spike identified in Figure \ref{AR11158_time_series}.}\label{AR11158_maps}
\end{figure}

The next example region that we present is much more complex than AR11318. AR11158 is a highly-studied active region due to it possessing both X-class flares and multiple CMEs \citep[e.g.][]{schrijver11,sun12,tziotziou13,inoue15, kay17, chintzoglou19, moraitis21, lee21}. Here, we will focus on the second CME from AR11158 listed in Table \ref{table}. The time series of $\dot{L}$ for the period before this CME is shown in Figure \ref{AR11158_time_series}.
Unlike AR11318, there are many flares in this active region. An X2.2-class flare, beginning at 19.73 hours on the time scale of Figure \ref{AR11158_time_series}, is associated with the CME and is preceded by the largest winding spike at 19.4 hours. The LASCO CME time corresponds to 20.4 hours and the ALMANAC time is approximately co-temporal with the X2.2 flare start time (a 4-minute difference). As before, the location of the winding signature can be seen by considering the maps in Figure \ref{AR11158_maps}.

There is a clear PIL displayed at the centre of the region in Figure \ref{AR11158_maps} (a) {at 35$^\circ$ Carrington longitude}. On the {top of the} eastern side of the PIL, there is a very strong change in $\mathcal{L}$, corresponding to the winding signature. {The global view of this feature is shown in Figure \ref{AR11158_maps} (b) and a zoomed-in section of the $B_z$ map overlaid with contours indicating the strongest part of  $\dot{\mathcal{L}}$ (displaying values of $\pm$60000 km$^2$s$^{-1}$) is shown in (c). }

\begin{figure}
    \centering
    \includegraphics[width=0.9\linewidth]{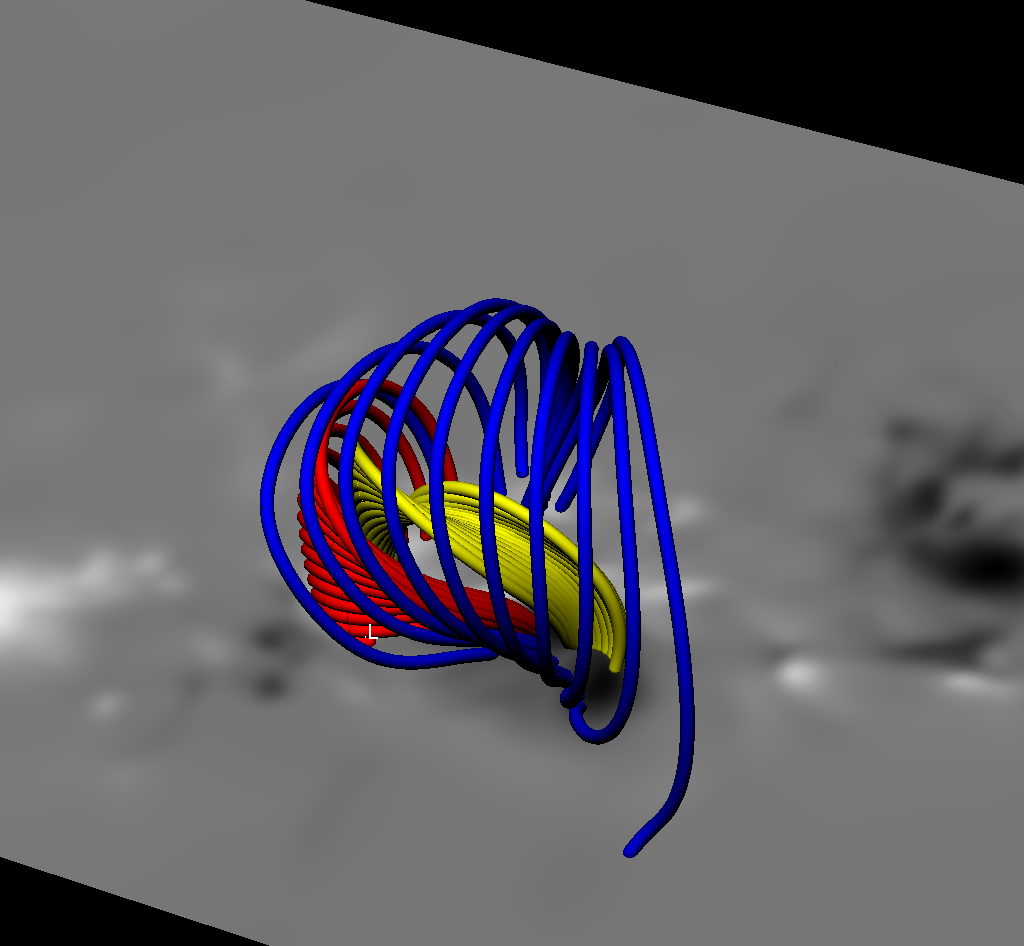}
    \caption{A nonlinear force-free extrapolation of the field lines of AR11158 at the time of the winding signature. The core of the flux rope is indicated by the yellow field lines. The red field lines, traced from the location of the winding signature, marked with ``L'', trace out one of the outer ``J''s of the flux rope. Blue field lines trace surrounding field. Note that the region is rotated by almost 180$^\circ$ with respect to the maps in Figure \ref{AR11158_maps}. The metrics for this extrapolation are $\lambda=0.0025$ and $\theta_J=6.96^\circ$.}
    \label{AR11158_field_lines}
\end{figure}

An approximation of the magnetic topology of the field lines connected to and near the winding signature is shown in Figure \ref{AR11158_field_lines}.
At the location of the winding signature, {marked in Figure \ref{AR11158_field_lines} by ``L'',} there is a strongly-twisted flux rope, whose main axis {lies across the central PIL. The core of the flux rope, extending high above the photosphere, is indicated by yellow field lines. The peak of the winding signature corresponds to field lines (shown in red) of the rope that form a highly twisted ``J'' structure - a common feature of sigmoidal flux ropes \citep[e.g.][]{titov99,hood12}. It is the change of this ``J'' structure, in contact with the photosphere, that leads to the winding signature.} The rope is surrounded by a magnetic arcade {(indicated by blue field lines)}. This configuration is the starting point for many eruption mechanisms (such as those cited in the Introduction) and matches that found in other extrapolations calculated for this region \citep[e.g.][]{inoue15,moraitis21}.

\subsection{AR11247}
\begin{figure*}
    \centering
    \includegraphics[scale=0.4]{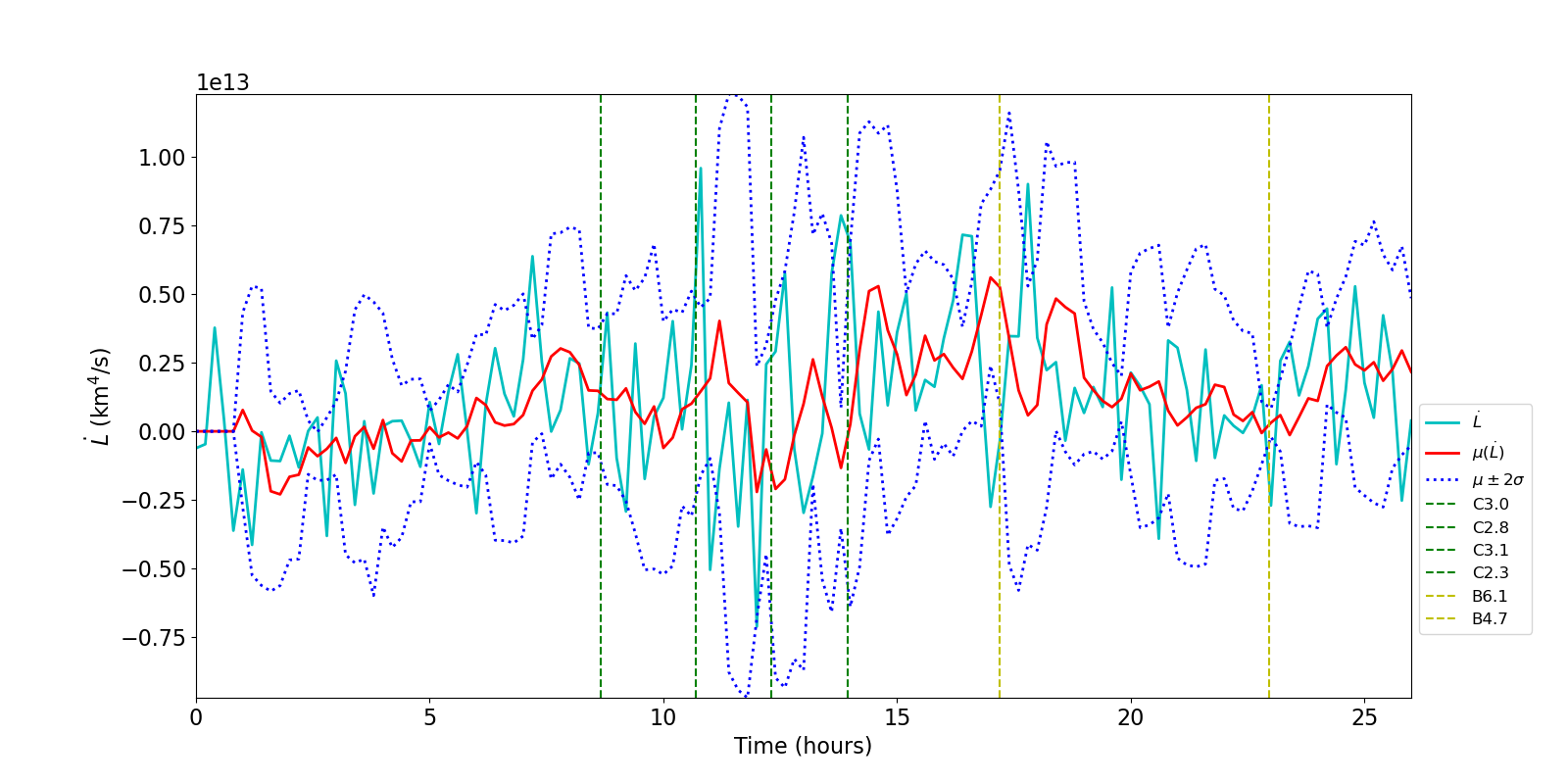}
    \caption{Time series of the magnetic winding flux rate $\dot{L}$ before the CME of AR11247. The quantities displayed follow the description in Part 1 of Section \ref{sec:sigs}. The LASCO CME time is at 23.8 hours on this graph.}
    \label{AR11247_time_series}
\end{figure*}

We now focus on the event of AR11247, which is the only example of the cases presented for which there is a large {(greater than 10$^\circ$)} separation in both longitude and latitude between the winding signature location and the CME location determined by \texttt{ALMANAC}. Although our approach does not seem to identify the coordinates (in space and time) of CME onset in this case, we nevertheless proceed as in previous cases in order to identify what is occurring in this event. As before, we first consider the time series of $\dot{L}$, which is shown in Figure \ref{AR11247_time_series}.

\begin{figure}
    \centering
    \begin{subfigure}{\columnwidth}
        \includegraphics[width=\textwidth]{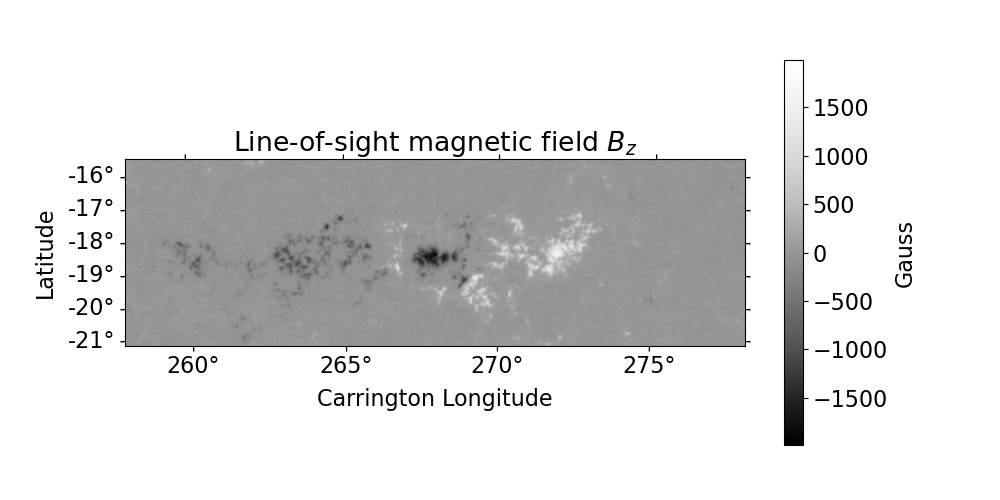}
        \caption{}
        \end{subfigure}

    \begin{subfigure}{\columnwidth}
        \includegraphics[width=\textwidth]{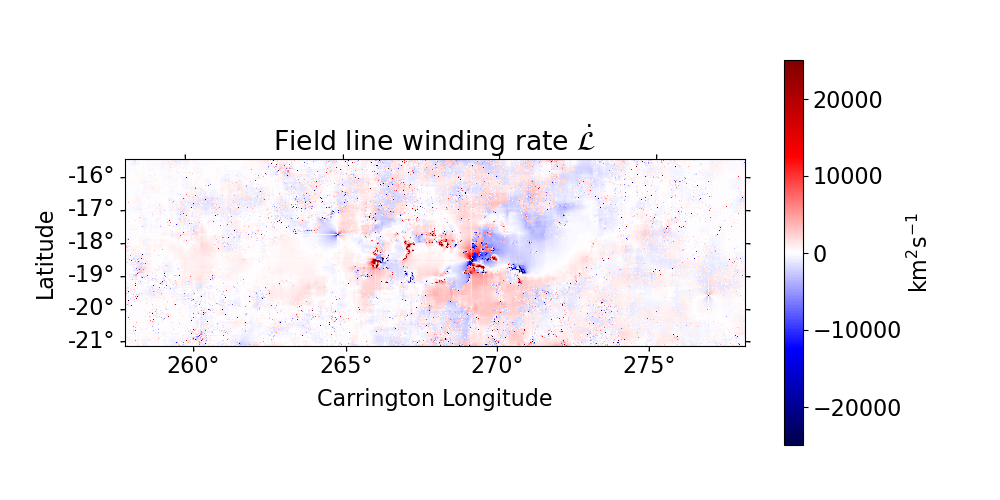}
        \caption{}
      \end{subfigure}

      \begin{subfigure}{\columnwidth}
        \includegraphics[width=\textwidth]{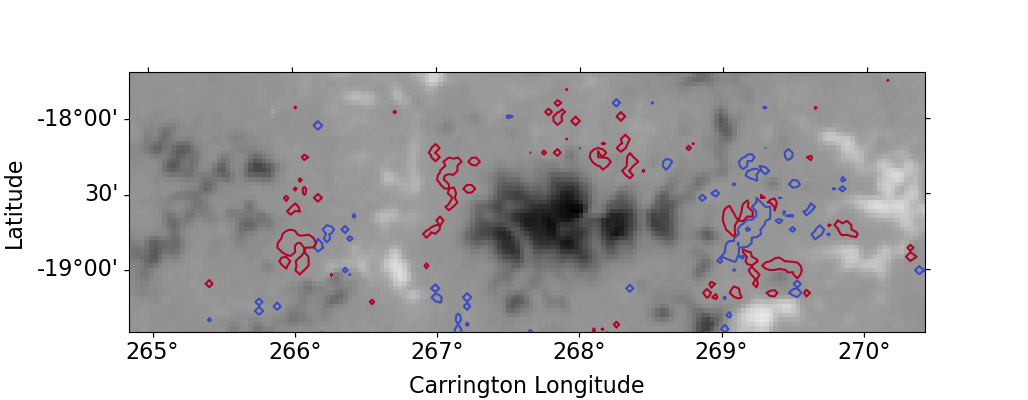}
        \caption{}
      \end{subfigure}
    \caption{(a) shows a map of $B_z$, (b) a map of $\dot{\mathcal{L}}$ and (c) a detail of the $B_z$ map with contours ($\pm10000$ km$^2$s$^{-1}$) of $\dot{\mathcal{L}}$ highlighting several possible locations for the winding signature. The strongest signature corresponds to the region at 269$^\circ$. All maps correspond to the time of the winding spike identified in Figure \ref{AR11247_time_series}.}\label{AR11247_maps}
\end{figure}

Given that the LASCO observation of the CME is at 23.8 hours on this figure, the B4.7 flare is the only possible flare that may be associated with the CME (if, indeed, the CME has an associated flare). However, the \texttt{ALMANAC} observation time is approximately 1 hour before the LASCO time, when there is no obvious winding signature. In order to understand the significance of these results better, we proceed with our previous approach and investigate the winding signature co-temporal with the start of the B4.7 flare, in order to see what connection, if any, it has with the CME. The maps of $B_z$ and $\dot{\mathcal{L}}$ at the time of the winding signature are displayed in Figure \ref{AR11247_maps}.

\begin{figure}
    \centering
    \includegraphics[width=0.9\linewidth]{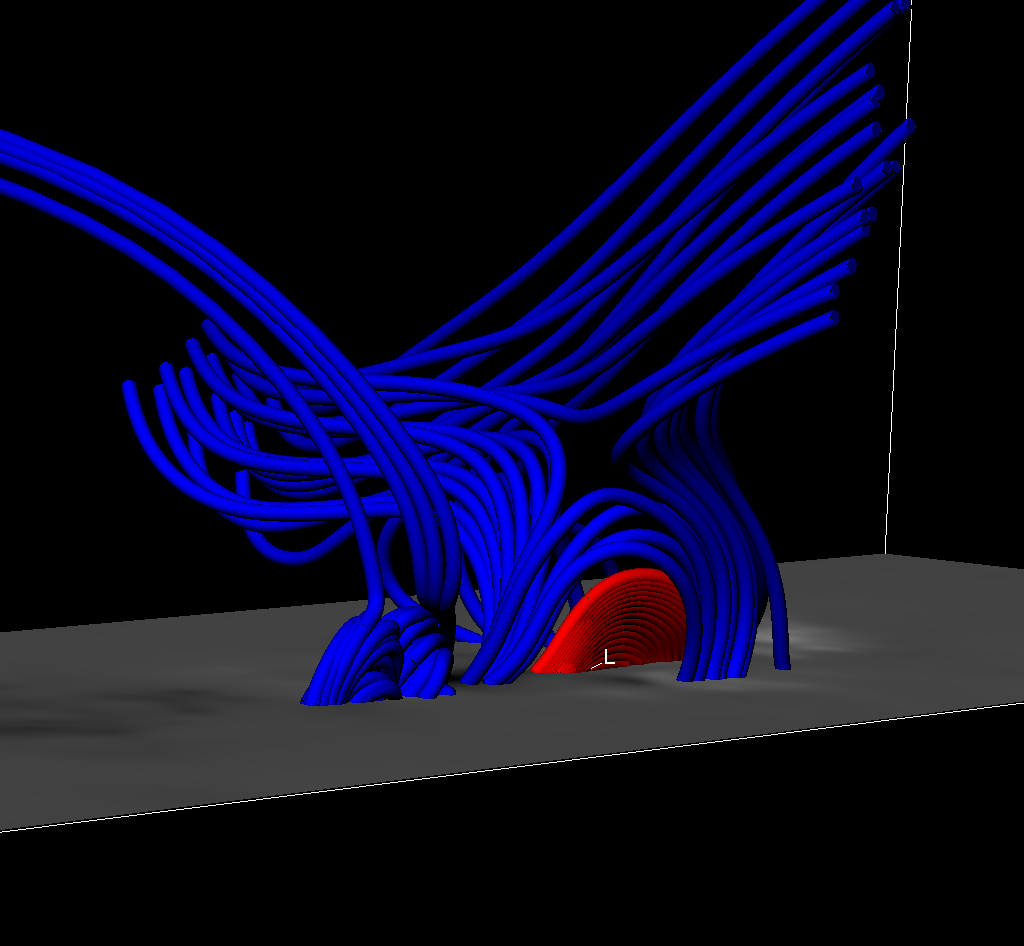}
    \caption{A nonlinear force-free extrapolation of the field lines of AR11247 at the time of the winding signature. The field lines, shown in red, related to the possible winding signature are traced from the position marked ``L''. Surrounding topological features are revealed by blue field lines. The metrics for this extrapolation are $\lambda=0.0042$ and $\theta_J=18.19^\circ$.}
    \label{AR11247_field_lines}
\end{figure}

{We follow our previous approach and select the strongest localized concentration of magnetic field line winding flux, located just above a small PIL south-east of the central negative polarity at 269$^\circ$ Carrington longitude}. As indicated in Table \ref{table}, the position of the CME is much further west and south of the winding signature. A possible connection, however, can be found by considering a force-free extrapolation, as shown in Figure \ref{AR11247_field_lines}.

\begin{figure}
    \centering
    \includegraphics[width=\columnwidth]{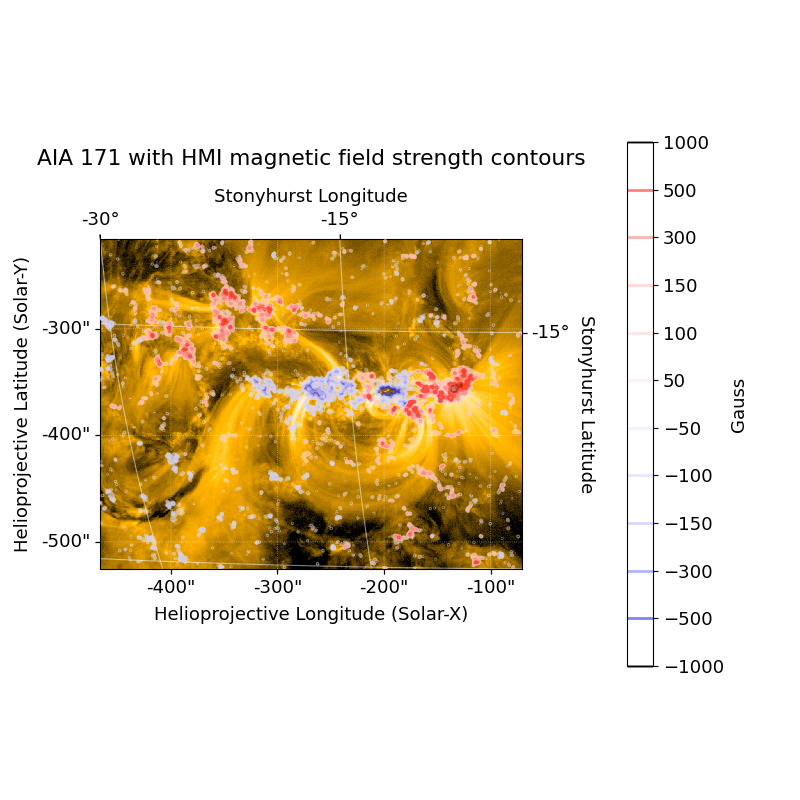}
    \caption{AIA 171 \AA~with line-of-sight magnetic field $B_z$ superimposed of AR11247 at the time of the winding signature from Table \ref{table}.}
    \label{aia}
\end{figure}

{The location of the selected winding signature corresponds to a region beneath the null point of a quadrupolar magnetic field.} Although this is the classical setup of the breakout model for CMEs \citep{antiochos99}, as mentioned previously, the CME does not occur here. Near to this region,  the magnetic field has a topology common to magnetic jets \citep{shibata07,pariat15}. The field lines higher in the domain are open, in the sense that they do not connect back down to the lower boundary. This suggests that there may be a connection between the region of the winding signature and magnetic fields outside this domain. Pursuing this line of enquiry, Figure \ref{aia} displays the line-of-sight magnetic field superimposed on AIA 171 \AA~ data at the time of the winding signature.

In the south-western corner of the map in Figure \ref{aia}, the dark arc, with a helioprojective longitude of -400'', is the rising filament of the CME. Based on the AIA image, the blue field lines from Figure \ref{AR11247_field_lines} connect to closed loops and not to the CME directly. One possibility, therefore, is that the expansion of the CME acted as a perturbation on the magnetic loops of AR11247, resulting in a localised change in magnetic topology, leading to the winding signature and the weak B-class flare (the \texttt{ALMANAC} time precedes the B4.7 flare time, see Tables \ref{table} and \ref{table_flare}). 

This analysis shows that this particular CME is connected to AR11247 but is not ejected from one of its internal PILs (even though it has been previously catalogued as emanating from this region \citep{pal18}). Thus, our approach can help to confirm whether or not a CME does originate from a particular active region. Further, such as in this case, our approach can still be used to analyse the topological complexity of an active region.

\subsection{Regions with ``strong'' penumbrae}
\begin{figure}
    \centering
    \begin{subfigure}[b]{\linewidth}
        \includegraphics[width=\columnwidth]{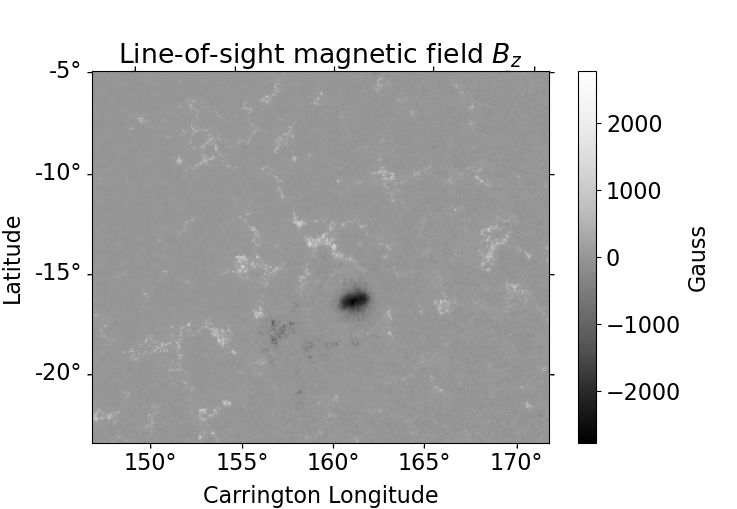}
        \caption{}
        \end{subfigure}

    \begin{subfigure}[b]{\linewidth}
        \includegraphics[width=\columnwidth]{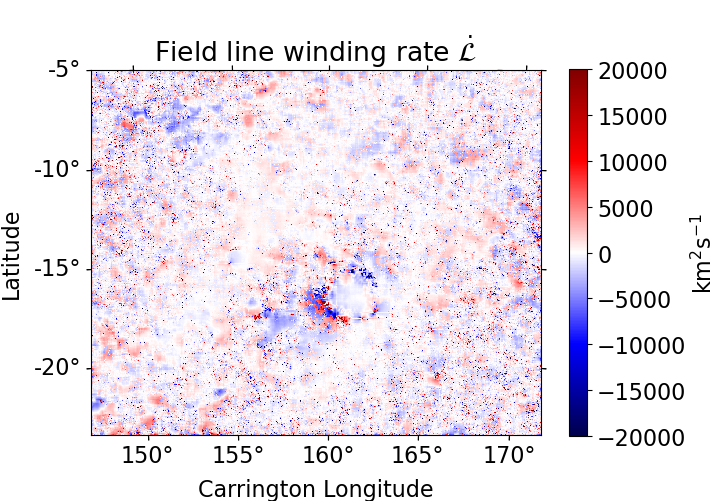}
        \caption{}
      \end{subfigure}
    \caption{Maps of (a) $B_z$ and (b) $\dot{\mathcal{L}}$ during the evolution of AR11777.}\label{penumbra}
\end{figure}

The approach we have outlined in this work can be applied to any SHARP region. For some regions, however, winding signatures at PILs can be masked if there is a particularly dominant sunspot with a developed penumbra. Since a penumbra is a region of magnetoconvection with a near-horizontal magnetic field, magnetic winding is particularly sensitive to field line motions in penumbrae. The result is that the largest spikes in time series of $\dot{L}$ are dominated by penumbra dynamics and, unless the CME and/or associated flare is very strong, the signatures at the PIL are weaker in comparison and not easily detected as a winding signature. An example of a winding signature dominated by a penumbra, from AR11777, is shown in Figure \ref{penumbra}.

For active regions such as these, in order to apply the approach of this article, the penumbra would have to be masked. This task, however, goes beyond the scope of this work.

\section{Summary and conclusions}
In this work, we present evidence that changes in magnetic topology at the photosphere are a common signature of CME onset. In particular, we show, by means of magnetic winding flux, that strong topological changes in the photosphere, occurring at a polarity inversion line within the active region, typically precede or are approximately co-temporal with early sightings of CMEs. We have found this result by analysing time series and maps of magnetic winding flux in 20-hour periods before the recorded coronograph observations (from LASCO or STEREO) for 30 CME events. From the time series, the times of winding signatures are identified, and their locations {are found by inspection of where the strongest concentrations occur in the}  field line winding maps. All data related to winding are based on SHARP magnetograms and processed with the \texttt{ARTop} code. In order to provide further evidence that the winding signatures are associated with CME onset, we also searched for the CMEs using the \texttt{ALMANAC} code. This makes use of AIA data and, thus, provides an earlier observation of a CME than LASCO or STEREO. For 28 out of the 30 events studied, there is {a good} match between the locations of the winding signatures and the CME locations determined by \texttt{ALMANAC}, {with differences being less than both the longitudinal and latitudinal extents of the active region at the photosphere}. The winding signatures either precede or are approximately co-temporal with the \texttt{ALMANAC} times, which all precede the recorded LASCO or STEREO observation times. For the remaining two events, data from \texttt{ALMANAC} is unavailable for one, and the other was not directly linked to the region for which it had been catalogued under.

The start times of the associated flares of each of the CMEs has also been compared to the times of the winding signatures. For all CMEs associated with flares of class C and above, the winding signatures precede or are approximately co-temporal with the flare start times.

Our main result is the following. For many active regions, magnetic winding flux can be used to identify, both spatially and temporally, the location of CME onset (and, if present, the associated flare-onset) within an active region. We provide a method to determine these properties that is based on the analysis of HMI and AIA data through the \texttt{ARTop} and \texttt{ALMANAC} codes respectively. Exceptions to this approach include regions for which penumbrae dominate the winding signatures. However, in the many cases for which the approach does work, and as demonstrated, the method allows for the identification of which PIL within the active region is directly related to the dynamics of CME onset. Detection of this location then allows for further investigation of CME onset.

In order to define an algorithm for detecting signatures of CME onset, we have given a precise definition of the winding signature in terms of the largest spike in $\dot{L}$ nearest to the recorded coronograph CME time. However, as is clear from the time series presented earlier, there are many other winding spikes and, often, they are associated with flares (but not CMEs). Although we have not focused on them in this work, all such spikes are potentially important in understanding the relationship between magnetic topology and flares, and can be studied in a similar manner to the way presented in this work.  

As well as identifying the coordinates of CME onset, our approach may also be useful in helping to determine the onset mechanism of a particular CME. The temporal relationship between the winding signature, an associated flare and early CME observations, puts a constraint on which particular mechanism may be at work during CME onset. This feature will be studied in future applications. 

\section*{Acknowledgements}

OPMA, DM and LF acknowledge support from the Leverhulme Trust, grant number RPG-2023-182. DM also acknowledges support from the Science and Technologies Facilities Council (STFC), grant number ST/Y001672/1 and LF acknowledges support from ST/X000990/1. \texttt{ARTop} was partly developed under an Innovation Placement supported by the DiRAC HPC Facility as part of the DiRAC Federation Project. Funding for the DiRAC Federation Project was provided by the UKRI Digital Research Infrastructure programme. \texttt{ALMANAC} was developed under Leverhulme grant RPG-2019-361. This research used version 5.1.0 of the SunPy open source software package \citep{sunpy_community2020}.

\section*{Data Availability}

Input files for the \texttt{ARTop} code for the 30 events presented here are available \href{https://researchdata.gla.ac.uk/1654/}{here}.

The \texttt{ARTop} code is available \href{https://github.com/DavidMacT/ARTop}{here} and the \texttt{ALMANAC} code is available \href{https://github.com/DrTomWilliams/ALMANAC}{here}.



\bibliographystyle{mnras}
\bibliography{phot_cme} 




\appendix

\section{Magnetic winding definition}
\label{appendix}
For detailed descriptions of magnetic winding, we guide the reader to \cite{prior20,mactaggart21gafd,mactaggart21natcomms}. Here, we just list the key formulae for completeness.

The calculation of magnetic helicity flux from magnetograms has become a standard calculation. For the calculation of magnetic helicity $H$, one formulation of the helicity flux, based on field line winding, is written as
\begin{equation}\label{helicity}
    \frac{\d H}{\d t} = -\frac{1}{2\pi}\int_P\int_P\frac{\d}{\d t}\theta(\x,\y)B_z(\x)B_z(\y)\,\d^2x\d^2y,
\end{equation}
where $P$ is the horizontal plane of the photosphere (the magnetogram), $\x$ and $\y$ are locations of field lines intersecting $P$ and $\theta$ is the pairwise winding angle of $\x-\y$. The first part of the  {integral} in equation (\ref{helicity}), involving the rate of change of the winding angle $\theta$, encodes topological changes in the magnetic field at the photosphere. This part is weighted by the magnetic flux, e.g. at $\x$ there is a local flux of $B_z(\x)\,\d^2x$.

Renormalizing equation (\ref{helicity}) to remove the flux weighting, produces a direct measure of the field line topology, called magnetic winding,
\begin{equation}\label{winding}
    \frac{\d L}{\d t} = -\frac{1}{2\pi}\int_P\int_P\frac{\d}{\d t}\theta(\x,\y)\sigma_z(\x)\sigma_z(\y)\,\d^2x\d^2y,
\end{equation}
where $\sigma_z(\x)$ is an indicator function taking the sign of $B_z(\x)$, and, therefore, can be -1, 0 or 1.

Equations (\ref{helicity}) and (\ref{winding}) represent the total  {(spatially-integrated)} changes in helicity and winding, respectively, over $P$. In this work, equation (\ref{winding}) is used to create time series. In order to gain spatial information about where strong changes in winding are occurring in an active region, we remove the outer integral from equation (\ref{winding}) to consider the field line winding rate
\begin{equation}
    \frac{\d}{\d t} \mathcal{L}(\x) = -\frac{\sigma_z(\x)}{2\pi}\int_P\frac{\d}{\d t}\theta(\x,\y)\sigma_z(\y)\,\d^2y.
\end{equation}
This quantity provides the winding rate of the field line at $\x$ relative to all the other field lines, and it is in this sense that we refer to $\mathcal{L}$ as the field line winding.


\bsp	
\label{lastpage}
\end{document}